# Angles in the SI: a detailed proposal for solving the problem

Paul Quincey
National Physical Laboratory[1], Hampton Road, Teddington, TW11 0LW, United Kingdom.
paulqquincey@gmail.com

**Abstract**

A recent Letter [1] proposed changing the dimensionless status of the radian and steradian within the SI, while allowing the continued use of the convention to set the angle 1 radian equal to the number 1 within equations, providing this is done explicitly. This would bring the advantages of a physics-based, consistent, and logically-robust unit system, with unambiguous units for all physical quantities, for the first time, while any upheaval to familiar equations and routine practice would be minimised. More details of this proposal are given here. The only notable changes for typical end-users would be: improved units for the quantities torque, angular momentum and moment of inertia; a statement of the convention accompanying some familiar equations; and the use of different symbols for $\hbar$ the action and $\hbar$ the angular momentum, a small step forward for quantum physics. Some features of the proposal are already established practice for quantities involving the steradian such as radiant intensity and radiance.

**Background to the current position**

The CCU[2] document 80-6 from 1980 has recently been drawn to my attention. This unpublished document[3] correctly points out that angles can be treated as base units, and that this would have the effect of introducing a dimensional constant into many equations. The constant was called $\varphi_0$, and is the same as $\theta_N$ in [1]. This constant is an angle with the value of exactly 1 radian, or about 57.3°. As this has been named the Cotes[4] angle [2], it is appropriate to rename the constant as $\theta_C$, which is the symbol that will be used here. In 1980 the proposal to give angles base unit status in the SI was turned down, though with a notable lack of full agreement among the members of the CCU, on the grounds that the advantages would be outweighed by the upheaval caused to current practice. The President of the CCU gave a short personal summary of the arguments [3]. I used similar reasoning to reach the same conclusion a few years ago [4][5].

Two points have recently become apparent which have led to me changing my mind, and which are good reasons why this topic should be reconsidered by the CCU. Firstly, it has been clearly demonstrated that the arguments used to support the idea that angles are inherently dimensionless are flawed [6], and that angles should instead be seen as inherently dimensional, on simple physical grounds relating to the conservation of angular momentum [2]. This explains why treating angles as inherently dimensionless has always led to anomalies and problems – in the distinction between frequency and angular frequency[6], and associated disappearing factors of 2π; in the question of whether 'rad' should be included in the units for angular quantities; in dimensional analysis; and in software, for example. It also explains why the decision made in 1980 evidently has not settled the matter. The choice is not simply a matter of personal preference and international diplomacy. Closing down discussion and sticking with the decision from 1980 is, in effect, just kicking the can down the road.

---

[1] This paper expresses the views of the author and these do not necessarily reflect those of NPL Management Ltd.
[2] The Consultative Committee for Units (CCU) advises the International Committee for Weights and Measures (CIPM), which in turn promotes world-wide uniformity of measurements units.
[3] The document was a note prepared by the Secretariat of ISO/TC 12 (the Technical Committee for Quantities and Units, which is responsible for the ISO 80000 standards) in collaboration with its Advisory Panel.
[4] Roger Cotes (1682-1716), a mathematician who worked with Isaac Newton, appears to have been the first person to realise the significance of this angle as a natural unit, in work published posthumously in 1722. Much earlier Indian mathematicians, such as Lalla in the 8[th] century, deserve special mention for using an accurate value of the angle in a related context. Though expressed as an angle (3438' = 57°18'), it seems that the value was used to set the radius of a drawn circle so that it had a convenient circumference, for work in trigonometry; it was not used as an angle unit [5].
[5] Note that in reference [4] the symbol $\eta$ is identical to $1/\theta_C$, where $\theta_C$ = 1 rad, as defined here. Of course, setting $\theta_C$ equal to the number 1 is the same thing as setting $\eta$ equal to the number 1.
[6] The proposal here to give frequency the unit s$^{-1}$ and angular frequency the unit rad/s differs from the current SI, where both have the unit s$^{-1}$, and also other proposals, such as [7], where both have the unit rad/s.

Secondly, as explained in the Letter [1], any upheaval associated with the change in status can be minimised by realising that the need to *define* angles as dimensional is separable from the question of whether or not to *treat* angles as dimensional within equations. The current position (as explained in [2] and [6]) is that there is a longstanding, widespread convention[7] to set $\theta_C$ equal to 1 within equations, without saying so explicitly, so that both $\theta_C$ and the inherent dimensionality of angle have been well hidden. This has been called the radian convention [2]. The situation has meant that the equation 1 rad = 1 (and, therefore, 1° ≈ 0.01745) is widely considered to be a universal truth, which is more than a little strange when pointed out directly. It is in fact just a convenient convention for simplifying equations, as any '$\theta_C$'s that should be in them can be crossed out.

When angles are correctly defined as having an independent dimension, this convention can still be adopted, whenever desired, but this must now be done explicitly with a statement such as 'where $\theta_C$, which has a value of 1 rad, has been set equal to 1 (so that the angle unit must be the radian, and the units on either side of the equation do not match)'. In this way familiar equations, including those for trigonometrical functions, can be left unchanged, but everyone is made aware that such equations are not the general, unit-invariant, i.e. 'complete' [2] equations. The statement would also highlight the fact that complete equations, within which the dimensions balance and the units match, are always available, and these are the same as the familiar equations except that the crossed-out '$\theta_C$'s have been reinstated. This logically-robust position would surely be a significant step forward from where we are now. The official position could be summarised as "angles are different; get used to it", when it should be "angles are no different from other quantities, but they are traditionally treated differently within equations". Examples of complete equations that differ from their familiar versions are given below.

It may be helpful to realise that this practice of crossing out a constant for convenience is not unique to $\theta_C$. The equation 1 rad = 1 is directly analogous to the equation 299 792 458 m/s = 1, one of the conventions adopted when 'natural units' are used, i.e. 'where *c* has been set equal to 1'. This is convenient for presenting simplified equations in mathematical physics – it allows the '*c*'s to be crossed out - but it is not an integral part of the mathematics or physics. Velocity is not inherently dimensionless. Moreover, when numerical calculations are performed, for example within software, the '*c*'s need to be put back, but you cannot be sure from the simplified equation on its own where they go. The decision to set *c* equal to 1 is always made explicit, not just to remind you to put the '*c*'s back when doing a calculation, but also to remind you that you need extra information to tell you where they go.

The definitions of trigonometric functions such as sin $\theta$ and cos $\theta$ need care when angle units other than radians are considered. The functions are commonly used in two distinct ways: sin x, where x is dimensionless, and the familiar expansion sin x = x - ⅙x³ +… applies, which is the convention throughout mathematical physics; and sin $\theta$, where $\theta$ is an angle (in any angle unit), and the expansion is sin $\theta$ = $\theta/\theta_C$ - ⅙($\theta/\theta_C$)³ +…., which is the case when we write sin 30° = 0.5. Brownstein [8] gave a good description of the situation and suggested using a capital letter to distinguish the two functions: sin x and Sin $\theta$, such that Sin $\theta$ = sin ($\theta/\theta_C$). Of course, the radian convention removes the distinction between the two, which is why it is rarely pointed out.

A phase angle $\Phi$ generally describes the argument of a complex number. As an angle, a phase angle can be expressed in any angle unit, just like 'geometric' angles such as those measured with a protractor. However, whenever a complex number is written as $r\,e^{i\Phi}$, or $r$ exp(i$\Phi$), equal to $r$ (cos $\Phi$ + i sin $\Phi$), the phase angle $\Phi$ must be in radians or the standard exponential function must be modified. The expression when any angle unit is allowed is $r$ exp(i$\Phi/\theta_C$), equal to $r$ (Cos $\Phi$ + i Sin $\Phi$), using Brownstein's notation – in other words, the familiar expression has simply adopted the radian convention. The complete expressions make it clear that the exp, cos and sin functions always operate on dimensionless quantities, but the radian convention makes it appear that the phase angle itself must be dimensionless. The ubiquitous adoption of the radian convention for mathematical descriptions of wave phenomena, for example, means that it is tempting to treat phase angle (or 'phase') as inherently dimensionless, and therefore different from a dimensional 'geometric' angle. It would be grossly inconsistent, though, to say that the product $\omega t$, where $\omega$ is an angular frequency, is an angle, whereas $\omega t$ within the expression $e^{i\omega t}$ is dimensionless. The situation is better seen as a strong argument for recognizing that the familiar 'radian convention' equations will continue to be used, while pointing out that these equations have been simplified from the complete, unit-invariant equations. As before, simplified equations should not be used to infer that angle (whether geometric or phase) is dimensionless.

---

[7] The word 'convention' is used in the sense of a choice that is agreed by many people but which is fundamentally optional; not a binding legal agreement like the Metre Convention. The radian convention was probably first adopted by Leonhard Euler in his work on the motion of rigid bodies in about 1765.

Unlike the redefinition of the kilogram, which was an idea whose time had come, the angle situation is a mistake that needs to be corrected. The prevailing attitudes until recently have been either that 1 rad = 1 is a universal truth, and so, whatever anomalies arise, there should be no correction; or that the correction to the status of angles should simply be made, and any consequent issues can be worked around or sorted out later; or that the correction is not absolutely necessary and the level of upheaval involved makes it undesirable. However, the proposal given in outline in the Letter [1] would be a logically-robust, workable, long-term solution to this genuine problem that has been habitually avoided. A more complete description of the proposal is presented below.

**The specific proposed changes (in bold), with clarifying information**

Changes to the status and definition of the radian and steradian within the SI

The radian is currently a dimensionless derived unit within the SI, equal to 1 m/m, while the steradian is a dimensionless derived unit equal to 1 $m^2/m^2$. **The radian would become either a new base unit or a 'complementary' unit** (the choice of term is a matter of preference and diplomacy), defined independently of the current base units. **The steradian would become a derived unit equal to 1 $rad^2$.** The dimensional equivalence of solid angle and (plane angle)$^2$, analogous to area and (length)$^2$, is most apparent in the limit of small angles, but is still true in the general case [8-10].

**The radian would need a new definition, such as: 'The SI unit of plane angle is the radian. The plane angle at a corner of a square, in other words one quarter of a complete rotation around a point in a plane, is exactly π/2 radians.'**

This definition avoids any lingering implication that angle is fundamentally a length ratio. The definition could be phrased in terms of the Cotes angle being given a value of exactly 1 radian, but this could seem too convoluted. The Cotes angle being an unfamiliar term, and a mathematical rather than a physical or technical constant, could be reasons for the definition having a different form from those for the current set of base units.

As a coherent derived unit, the steradian does not need its own definition, but the unfamiliar nature of spherical trigonometry [10] means that it is helpful to explain how large a steradian is, independently of the radian. This can be done by stating that 'the solid angle at a corner of a cube, in other words one eighth of the full field of view from any point, is exactly π/2 steradians'. This avoids the implication that solid angle is fundamentally an area ratio.

The changes to status and definition are necessary to make the SI a logically-robust unit system. Of course, the size of the radian and steradian would remain exactly as they are at present, and these changes have no effect on the practical measurement of angles, solid angles, or related quantities, so the effect of the changes for end-users would be minimal. However, these changes on their own would not solve the existing problems, because they would appear to be incompatible with the fact that angles are treated as dimensionless within many equations, for example. The necessary accompanying changes, which have more noticeable consequences, though not major ones, are given below.

Changes to the units for certain angular quantities within the SI

The familiar units assigned to some angular quantities are based on equations that have adopted the radian convention, and so are missing 'rad's that would be present if the complete equation is used. The physically-correct units are those with the 'rad's reinstated. Numerical values would not change, and the physical meanings of all quantities would also be unaffected. **The SI units for:**

**Torque**[8] would change from N m (= J) to J rad$^{-1}$
**Angular momentum**[9] would change from J s to J s rad$^{-1}$ (i.e. J/(rad/s))
**Moment of inertia**[10] would change from kg m$^2$ to kg m$^2$ rad$^{-2}$ (i.e. J/(rad/s)$^2$)

**In addition, the option to omit the radian from the SI units for angle, angular velocity, angular frequency, angular acceleration, and angular wavenumber would be removed, the only correct SI units being rad, rad/s, rad/s, rad/s$^2$ and rad/m respectively.** The option has arisen from the contradiction between natural physical instincts, which imply that the radian should be included with the unit, and familiar equations (which have adopted the radian convention) which imply that it should not.

These proposals should not be thought of as new conventions. They are the physically-correct units, which are evident from complete equations, i.e. equations within which no simplifying convention has been adopted. The units should not revert back to the current versions whenever the radian convention is adopted, because specific quantities should have specific SI units, which should not change when conventions are used to simplify equations. The removal of options for the units for some angular quantities is, by itself, a significant advantage of this proposal, especially when software is involved.

It is illuminating to note that in the field of light and radiation, within which solid angles feature strongly in the definitions of quantities such as radiant intensity and radiance, the standard units unambiguously contain the angle unit – they are W sr$^{-1}$ and W sr$^{-1}$ m$^{-2}$ respectively. The steradian is not optional, even though 1 steradian is officially equal to the number 1, because this would lead to confusion between the lumen and the candela, for example, where 1 cd = 1 lm sr$^{-1}$. So unlike torque, where the angle unit is officially absent, or angular velocity, where the angle unit is officially optional, for radiance the angle unit is officially compulsory. No serious unit system should be satisfied with this unnecessary lack of consistency. The units within the field of light and radiation effectively already treat angle as a base quantity, and would therefore not need to change if the status and definition of the steradian were changed in the way described here. To quote from the CCU 80-6 document: 'In this field…if it [angle] becomes a base quantity …there is a clear advantage'.

The clarity that an independent unit for angle brings to radiation measurement could easily be brought to mechanics, in distinguishing between the units for torque and energy, and those for action and angular momentum, for example. This would also bring all areas of physics within a consistent, improved framework, where it is compulsory to use the physically-correct unit for all quantities.

Arguably, the role of the SI, and the scope of a potential CGPM[11] resolution, end at this point. However, the fact that many familiar equations apparently support the present dimensionless status of angles, and imply that the radian is not needed in the units for angular quantities, is an issue that should be addressed in parallel, at least in the form of SI recommendations.

---

[8] The distinction between torque ***T*** and moment of force ***M*** is a matter of definition. Moment of force could be made formally distinct from torque by retaining the definition ***M*** = ***r*** X ***F***, with the familiar units N m, while the equation for torque changes, as described in Table 2 below.

[9] Angular impulse, which is the product of torque and time, is equal to the change in angular momentum. Its units would therefore change from N m s to J s rad$^{-1}$, the same as for angular momentum.

[10] Although 1 sr = 1 rad$^2$, it is not helpful to use these units interchangeably. The steradian should be reserved for solid angles. This is in direct analogy with length and area. If there was an official, named SI unit for the area 1 m$^2$ – we just have an unofficial one, the centiare (ca) – we would only use it when an area is involved. We would still say 1 J = 1 kg m$^2$ s$^{-2}$, not 1 J = 1 kg ca s$^{-2}$, because the m$^2$ is part of a squared velocity; there is no relevant area. Similarly, moment of inertia should have the unit kg m$^2$ rad$^{-2}$ (i.e. J/(rad/s)$^2$), not kg m$^2$ sr$^{-1}$.

[11] The General Conference on Weights and Measures, the highest decision-making body in global metrology.

Equations

The proposal that the radian convention could still be adopted within equations after angles are changed in status, when this is done explicitly, means that all the familiar equations could still be used. However, **where they are different from the complete version, equations should be accompanied by a statement such as 'where $\theta_C$, which has a value of 1 rad, has been set equal to 1 (so that the angle unit must be the radian, and the units on either side of the equation do not match)'.** This would be the only visible change. Of course, if only complete equations are used, the statement would not be needed.

The complete equations, which are valid for any angle unit and which are dimensionally balanced, so that the units match, should be available in some convenient place, if only for use in software and educational purposes. In *most* cases, the rule is that where familiar equations contain an angle $\theta$, an angular velocity or angular frequency $\omega$, an angular wavenumber $k$, or a solid angle $\Omega$, these must simply be replaced by $\theta/\theta_C$, $\omega/\theta_C$, $k/\theta_C$ and $\Omega/\theta_C^2$ respectively to form the complete equation. The relationship $\omega = d\theta/dt$ follows this rule, and is the same in both versions as the '$\theta_C$'s cancel. Examples where the rule is applied and the two versions are different are given in Table 1.

| Quantity | Familiar equation | Complete equation |
| --- | --- | --- |
| angular frequency | $\omega = 2\pi f$ | $\omega = 2\pi\theta_C f$ |
| angular wavenumber | $k = 2\pi/\lambda$ | $k = 2\pi\theta_C/\lambda$ |
| arc length | $s = r\,\theta$ | $s = r\,\theta/\theta_C$ |
| sector area | $A = \tfrac{1}{2} r^2\,\theta$ | $A = \tfrac{1}{2} r^2\,\theta/\theta_C$ |
| tangential velocity | $v = r\,\omega$ | $v = r\,\omega/\theta_C$ |
| centripetal acceleration | $a = r\,\omega^2$ | $a = r\,\omega^2/\theta_C^2$ |
| solid angle | $\Omega = A/r^2$ | $\Omega = \theta_C^2 A/r^2$ |

Table 1: Some familiar equations that are converted to complete equations following a simple rule for $\theta$, $\omega$, $k$ and $\Omega$.

To give the example of finding the period of oscillation of a mass $m$ attached to a spring of stiffness $k$, with the equation $m\,d^2x/dt^2 = -kx$, the approach using familiar equations would be to try a solution of the form $x = A \exp(i\omega t)$. This gives a result for angular frequency $\omega$ (with SI unit s$^{-1}$) of $\omega = (k/m)^{1/2}$. The period $P$ is given by $P = 2\pi/\omega$, so that the period of oscillation $P = 2\pi/(k/m)^{1/2}$ (with SI unit s).

To obtain the solution using complete equations we simply apply the rules above, trying a solution of the form $x = A \exp(i\omega t/\theta_C)$ to give $\omega = \theta_C(k/m)^{1/2}$, where $\omega$ now has the SI unit rad/s. Period $P$ is now given by $P = 2\pi\theta_C/\omega$, so that the period of oscillation is again given by $P = 2\pi/(k/m)^{1/2}$ (with SI unit s).

Slightly more thought is needed when solving the problem of a damped oscillation represented by the equation $m\,d^2x/dt^2 + \beta\,dx/dt + kx = 0$. The equation can be solved mathematically, i.e. before considering units, by trying a solution of the form $x = A \exp(\alpha t)$, where $\alpha$ is a complex number. In the familiar case we then interpret the results for $\alpha$ physically as $\alpha = -\gamma + i\omega$, where $\gamma$ and $\omega$ are the exponential decay rate and the angular frequency respectively. For the complete case, we use the same trial solution, producing the same solutions for $\alpha$, but, following the rules above, interpret the results using $\alpha = -\gamma + i\omega/\theta_C$. As before, the exp function must operate on a dimensionless quantity, so $\alpha t$ must be dimensionless, and in the familiar case both $\gamma$ and $\omega$ have SI units of s$^{-1}$, while in the complete case $\gamma$ and $\omega$ have SI units of s$^{-1}$ and rad/s respectively.

These examples are intended to show that no logical contradictions arise from the use of the complete equations involving angular quantities. It is not suggested that the complete equations should be used instead of the simpler familiar equations for routine problem solving.

Familiar equations containing the angular quantities torque $T$, angular momentum $L$, or moment of inertia $I$ can be covered by extending the rule, so that in complete equations they are replaced with $T\theta_C$, $L\theta_C$ and $I\theta_C^2$ respectively. They are treated separately here because these changes are less intuitive than those to the more basic angular quantities $\theta$, $\omega$, $k$ and $\Omega$, and it is helpful to show more of the equations that remain unchanged. Other equations treated separately are, firstly, from the field of light and radiation, either where the equation is already complete or it includes another angular quantity such as radiance $L$ (which needs to be replaced by $L\theta_C^2$), or, secondly, where an ambiguous symbol such as $\hbar$ is involved (as described in the next section). Examples are given in Table 2.

| Quantity | Familiar equation | Complete equation | |
|---|---|---|---|
| torque | $T = dL/dt = I\, d\omega/dt$ <br> $T = r \times F$ | $T = dL/dt = I\, d\omega/dt$ <br> $T = (r \times F)/\theta_C$ | unchanged |
| angular momentum | $L = I\omega$ <br> $L = r \times p$ | $L = I\omega$ <br> $L = (r \times p)/\theta_C$ | unchanged |
| moment of inertia | $I = \Sigma\, m_i r_i^2$ | $I = (\Sigma\, m_i r_i^2)/\theta_C^2$ | |
| rotational kinetic energy | $E = \tfrac{1}{2} I \omega^2$ | $E = \tfrac{1}{2} I \omega^2$ | unchanged |
| radiant intensity | $I_e = d\Phi_e/d\Omega$ | $I_e = d\Phi_e/d\Omega$ | unchanged |
| radiance of a black body | $L = \sigma T^4/\pi$ | $L = \sigma T^4/\pi\theta_C^2$ | |
| photon energy | $E = h\nu = \hbar\omega$ | $E = h\nu = \hbar\omega$ | unchanged |

Table 2: Some familiar equations that are not converted to complete equations following the simple rule for $\theta$, $\omega$, $k$ and $\Omega$.

It is helpful to clarify the status of the symbol π within these equations. In complete equations, π always represents the dimensionless number which is approximately 3.14159. In familiar equations such as $\omega = 2\pi f$, it is tempting to interpret the π as representing the angle π rad, but this leads to inconsistency, because in other familiar equations like $A = \pi r^2$, for the area of a circle, π represents a dimensionless number. The recommendation to add the statement "where $\theta_C$ has been set equal to 1…" to equations where the radian convention has been adopted brings full clarity and avoids the need for any 'rules of thumb' for π or anything else.

Clarification of some physical constants

As noted above, different physical quantities like torque and energy, with different physically-correct units, appear to have the same units when the radian convention is adopted. In the same way, two different specific physical quantities can appear to have the same numerical value and units, and this can lead to them being represented by the same symbol. By far the most prominent example is the reduced Planck constant $\hbar$, which is used to represent both an action and an angular momentum, when these are physically different quantities. As recommended in [2]:

- **$\hbar$ should be presented unambiguously as an angular momentum, $h/2\pi\theta_C$, with a value of about $1.05 \times 10^{-34}$ J s rad$^{-1}$, not an action.**
- **$\check{h}$ should be introduced as the action $h/2\pi$, with a value of about $1.05 \times 10^{-34}$ J s.**

One reason for assigning the existing symbol $\hbar$ to the angular momentum is that the familiar equations $E = \hbar\omega$ and $p = \hbar k$ are then also the complete equations. The use of different symbols for these two physically distinct quantities would be good practice and should be encouraged, even in equations where the radian convention is adopted, in which case both symbols effectively represent the same quantity (about $1.05 \times 10^{-34}$ J s). The symbol $\hbar$ in the expression for the spin of an electron, $\hbar/2$, and in the equation $E = \hbar\omega$, is an angular momentum, whereas the symbol $\hbar$ in the familiar expression for the fine structure constant, $\alpha = e^2/4\pi\varepsilon_0 \hbar c$, is an action that would be better represented by the symbol $\check{h}$. The equation would perhaps be better written as $\alpha = e^2/2\varepsilon_0 h c$. Some further examples are given in Table 2 in [2]. Others include the quantum operators $\widehat{L_z}$ = - iℏ ($y\, \partial/\partial x - x\, \partial/\partial y$) (in Cartesian coordinates), $\widehat{L_z}$ = - iℏ̌ $\partial/\partial\theta$ (in cylindrical or spherical coordinates), and $\hat{\boldsymbol{p}}$ = - iℏ$\Delta$.

It should be stressed that there is no simple rule for knowing when ℏ should be replaced with ℏ̌, and this needs to be done by considering the physical dimensions within each equation. The Planck constant $h$ would be unaffected, remaining as an action with the defined value 6.62607015 x 10$^{-34}$ J s.

To reemphasise one of the central points of this proposal, it does not require engineers or physicists to insert '$\theta_C$'s into their working equations to make them complete. It would, however, be occasionally helpful for them to realise that non-complete equations cannot be used for dimensional analysis, that the changes to units for angular quantities complete the process of constructing a fully logical unit system, and that the symbol ℏ is currently used for two distinct physical quantities.

Dimensional analysis

**Plane (and phase) angle should have its own dimension, A. Solid angles would have the dimension A$^2$.** Dimensional analysis would, for the first time, be a valid and useful tool for all complete equations, including those containing angles. Moreover, the statement recommended within this proposal would highlight the basic point that if you have set a dimensional constant equal to 1 within an equation, you should not expect dimensional analysis to work. The only truly dimensionless quantities would be ratios and counts.

**Conclusion**

Various advantages of giving plane angles independent, dimensional status, rather than the dimensionless status that was formally adopted within the SI in 1995, have been set out many times over a long period [e.g. 8, 9, 11-14]. Such advantages are implicitly already recognised within the SI for quantities such as radiant intensity and radiance. However, it has not been widely understood that this change in status, on its own, leads to problems with the units currently assigned to some quantities, and a logical requirement to add a dimensional constant to many familiar equations. These problems can be avoided by the simple accompanying changes described here, with the result that the advantages can be seen to easily outweigh the associated minor disturbances.

There would no doubt be initial reluctance among some educators, scientists and engineers to changing the SI units given to torque, angular momentum and moment of inertia, on the grounds that the current units are embedded into common practice, and people found ways to work around their logical inconsistencies long ago. However, the physical clarity this would build into the SI should be recognised very quickly. The units would tell us that *torque x angle = energy*, and *angular momentum x angle = action*, for example, in the same way that they do for *force x distance = energy*, *linear momentum x distance = action*, and *radiant intensity x solid angle = radiant flux*. Dimensional analysis could be used to its full extent. Software involving angular quantities would be rationalised. Arguments about the correct units for frequency and angular frequency, and the meaning of the unit Hz, could be left behind. The explanation of these changes would be considerably easier and more rewarding than explaining how a kilogram-sized mass can be measured in terms of the Planck constant.

**Note added after submission**

Since this paper was written, Brian Leonard's long, thoughtful and well-researched paper on the topic, *Proposal for the dimensionally consistent treatment of angle and solid angle by the International System of Units* [14], has been accepted for publication. Although there is a great deal of common ground, it will probably be helpful to readers to highlight the main differences between his proposal and the one put forward here.

1) Using $\theta_C$ or rad for the angle constant in equations. Of course, $\theta_C$ = 1 rad, and it is technically correct that the unit symbol rad represents a quantity of angle. However, like many of the people referenced in [14], I find it very helpful to reserve unit symbols for the specification of quantities, such as 1.2 rad, and not to use unit symbols within equations. The symbol $\theta_C$, apart from looking like a term normally seen in equations, emphasises that the quantity is independent of the choice of angle unit.

2) <u>Allowing the continued use of familiar equations, providing the convention involved is made explicit.</u> As mentioned above, I do not rule out the option of including the $\theta_C$ (or rad) symbols wherever they are needed to make equations complete, in which case no "explicitness" statement is required. However, I do not think it would be viable to 'sell' this approach to the user community. Physicists and engineers will continue to write $v = r\omega$ for tangential velocity, cos ($kx - \omega t$) for wave motion, and so on. The point is that they should be aware that these equations are simplified, at the expense of unit-invariance and dimensional analysis, not that they should be forced to change what they write down, or that they should be told that simplified equations are wrong.

3) <u>Setting $\theta_C$ equal to 1.</u> Leonard states that "the physical quantity $\theta_C$ … cannot be set equal to 1." In a physical sense, this is absolutely correct. But it is also correct that mathematical physicists commonly set dimensional quantities like $c$ or $4\pi\varepsilon_0$ equal to 1, to simplify their equations, because symbols can then literally be crossed out. This is a valid description of the way the SI (and nearly everyone else) treats angles. It is equivalent to Leonard's description in terms of a conventional "nondimensionalization" of angles, but, in my view, simpler to present and understand.

4) <u>Changing the units for torque, angular momentum and moment of inertia.</u> This is a more subtle, but vitally important point.

   i) To take the example of moment of inertia $I$, Leonard proposes keeping the 'conventional definition' $I = \Sigma m_i r_i^2$, in which case the kinetic energy of a rotating body will be $E_k = \frac{1}{2} I\omega^2/\text{rad}^2$ (to use his terminology). However, if we instead use the equation $I = \Sigma m_i r_i^2/\theta_C^2$, the kinetic energy stays as $E_k = \frac{1}{2} I\omega^2$, directly corresponding with $E_k = \frac{1}{2} mv^2$ for linear motion. Similarly, he proposes keeping the 'conventional definitions' for torque of $\boldsymbol{T} = \boldsymbol{r} \times \boldsymbol{F}$ and angular momentum of $\boldsymbol{L} = \boldsymbol{r} \times \boldsymbol{p}$, which requires changing the more general (and equally familiar) relationships $\boldsymbol{T} = I\, d\boldsymbol{\omega}/dt$ and $\boldsymbol{L} = I\boldsymbol{\omega}$, so that they lose their correspondence with $\boldsymbol{F} = m\, d\boldsymbol{v}/dt$ and $\boldsymbol{p} = m\boldsymbol{v}$. The simple relationship energy = torque x angle, corresponding to energy = force x distance, would also be lost in Leonard's proposal. We can see that the 'conventional definitions' are not actually *the* definitions; they are instead simplified equations that incorporate the radian convention (or, in other words, the "nondimensionalization" of angle). There are other 'defining equations', such as $E_k = \frac{1}{2} I\omega^2$, $\boldsymbol{T} = I\, d\boldsymbol{\omega}/dt$ and $\boldsymbol{L} = I\boldsymbol{\omega}$, which give a better indication of how best to proceed.

   ii) Taking angular momentum in another context, the conjugate variables energy and time, momentum and distance, and angular momentum and angle (well-known from the Uncertainty Principle) all form a product which is an action (with SI units J s). Leonard is advocating that the simplicity of these relationships should be lost, by inserting a "rad" into the familiar equation $\Delta L \Delta\theta \geq h/4\pi$, in order that angular momentum can retain its familiar units of J s. It is surely better, instead, to accept that angular momentum should properly have units of J s/rad, with this "rad" having been traditionally omitted because of the convention to "nondimensionalize" angles.

   iii) To summarise, the physical definitions and interrelationships of torque, angular momentum and moment of inertia would not be changed by the proposal described here. Their units would simply change to the ones they should have when angle is treated as a dimensional quantity. As Leonard notes in his paper, there has been widespread agreement on this in the published literature, since James Brinsmade showed us the way in 1936 [11].

**References**

[1] Paul Quincey, Angles in the SI: treating the radian as an independent, unhidden unit does not require the redefinition of the term 'frequency' or the unit hertz, Metrologia **57** 053001 (2020).